\documentclass{PoS}

\usepackage{amsmath}
\usepackage{bm}
\usepackage{xspace}

\newcommand{\pt}{\ensuremath{p_t}\xspace}
\newcommand{\ttbar}{\ensuremath{t\bar{t}}\xspace}

\title{Search for heavy resonances decaying to top quarks}

\ShortTitle{Search for heavy resonances decaying to top quarks}

\author{\speaker{Roman Kogler} \\
        University of Hamburg\\
        E-mail: \email{roman.kogler@physik.uni-hamburg.de} \\
        	On behalf of the CMS Collaboration.\\
        }

\abstract{
In many models of physics beyond the Standard Model the coupling of new states to third generation quarks is enhanced. A review is presented of searches by the CMS collaboration for heavy particles decaying to final states involving top quarks. This includes searches for heavy gauge bosons and excited states. Several final states originating from the top quark decays are considered and the event reconstruction is optimised accordingly. 
The analyses presented use data collected with the CMS experiment during the year 2012 at the LHC, in proton-proton collisions at a centre-of-mass energy of 8 TeV. 
}

\FullConference{The European Physical Society Conference on High Energy Physics \\
		 18-24 July, 2013\\
		 Stockholm, Sweden}

\begin{document}

\section{Introduction}
\vspace{-0.2cm}
The discovery of a new boson by the ATLAS and CMS Collaborations~\cite{Aad:2012tfa, Chatrchyan:2012ufa}, with properties closely resembling those of the Standard Model (SM) Higgs boson~\cite{ATLAS-CONF-2013-034, CMS-PAS-HIG-13-016}, marks a great success of the SM. Assuming the new boson is indeed the SM Higgs boson, its mass of $M_{H}\simeq125.6$~GeV is consistent with electroweak precision data~\cite{Baak:2012kk} and the electroweak sector of the SM is complete. 
However, within the SM many questions are still unanswered. The exact nature of electroweak symmetry breaking and the generation of Yukawa couplings is still unknown and subject to numerous extensions of the SM. Also, the validity of the SM up to the Planck-scale would need a large amount of fine-tuning without additional contributions from new physics that cancel the divergent contributions from the top quark, vector bosons and the Higgs boson itself. Because of the top quarks' large mass, it gives the largest contribution to $M_H$. 
This effect is mitigated in numerous extensions of the SM. These include supersymmetry, extensions of the SM symmetry groups or additional spatial dimensions. In many of these models new particles appear which either couple strongly to the top quark, or have similar quantum numbers as the top quark to cancel the divergent terms in $M_H$ through additional loop corrections. 

While experimental searches for top-partners in SUSY and for fourth-generation vector-like quark models are discussed elsewhere~\cite{exo_top:EPS13}
this report focuses on experimental searches for non-SUSY extensions of the SM predicting resonances decaying to top quarks. These include, among others, excited top partners~\cite{Hassanain:2009at}, heavy $Z'$ and $W'$ gauge bosons decaying to top quarks~\cite{Appelquist:2000nn, Agashe:2006hk}, compositeness models~\cite{Kumar:2009vs} and top quark condensation~\cite{Hill:1991at}. 

Searches for resonances decaying to top quarks have been carried out previously at the Tevatron collider and at the LHC using $p\!p$ collision data at   a centre-of-mass energy of $\sqrt{s}=7$~TeV. Here the results from new searches are reported, using the full $\sqrt{s}=8$~TeV dataset recorded with the CMS detector~\cite{Chatrchyan:2008aa} in the year 2012 with an integrated luminosity of $19.6$~fb$^{-1}$. These searches improve the previous results considerably in sensitivity, leading to a higher reach in resonance mass in some of the benchmark models considered. 


\vspace{-0.2cm}
\section{Resonances in $\bm{t b}$ production}
\vspace{-0.2cm}
In many extensions of the SM charged massive gauge bosons $W'$ are expected to couple to third-generation quarks more strongly than to quarks of the first and second generation. 
A recent search by the CMS Collaboration in the $t b$ final state~\cite{B2G-12-010} utilizes the $t \to Wb$ and subsequent $W\to \ell \nu$ decay chain, where $\ell$ is an electron or muon which may also originate from the $W\to \tau \nu$ and $\tau \to \ell \nu$ decay.
In this analysis events are selected 
by requiring one lepton with high transverse momentum of $\pt>50$~GeV and at least two jets with $\pt > 120$~GeV and $\pt > 40$~GeV, 
one of which has to be $b$-tagged. 
The expected SM backgrounds are taken from Monte-Carlo (MC) simulations. After the event selection the expected number of events from SM processes and the observed events in data show good agreement to better than 2\%. 
The largest background contributions are from \ttbar production with 62\% and $W$+jet production with 28\%. 

\begin{figure}[t!]
\begin{center}
\includegraphics[width=0.5\textwidth]{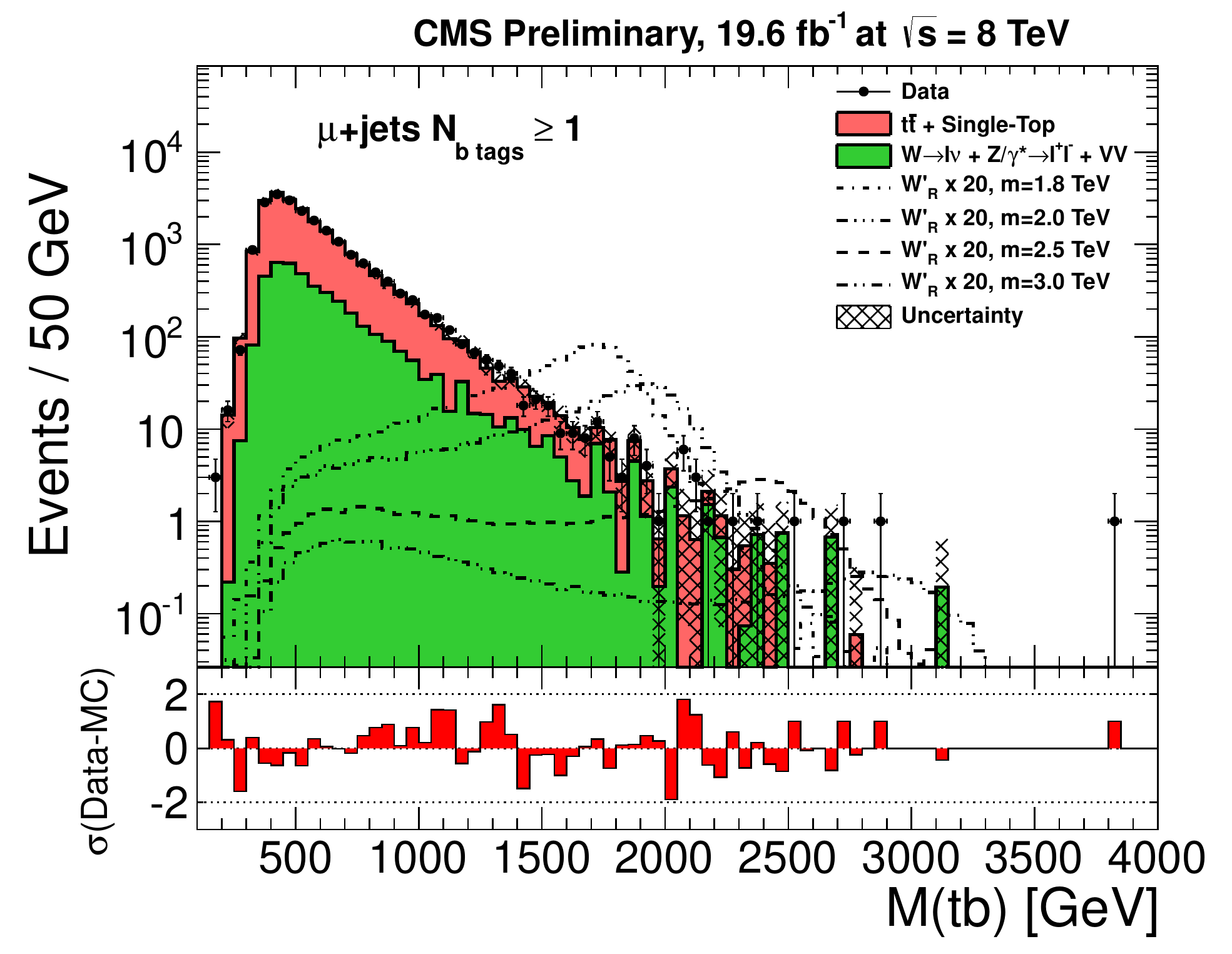} 
\includegraphics[width=0.415\textwidth]{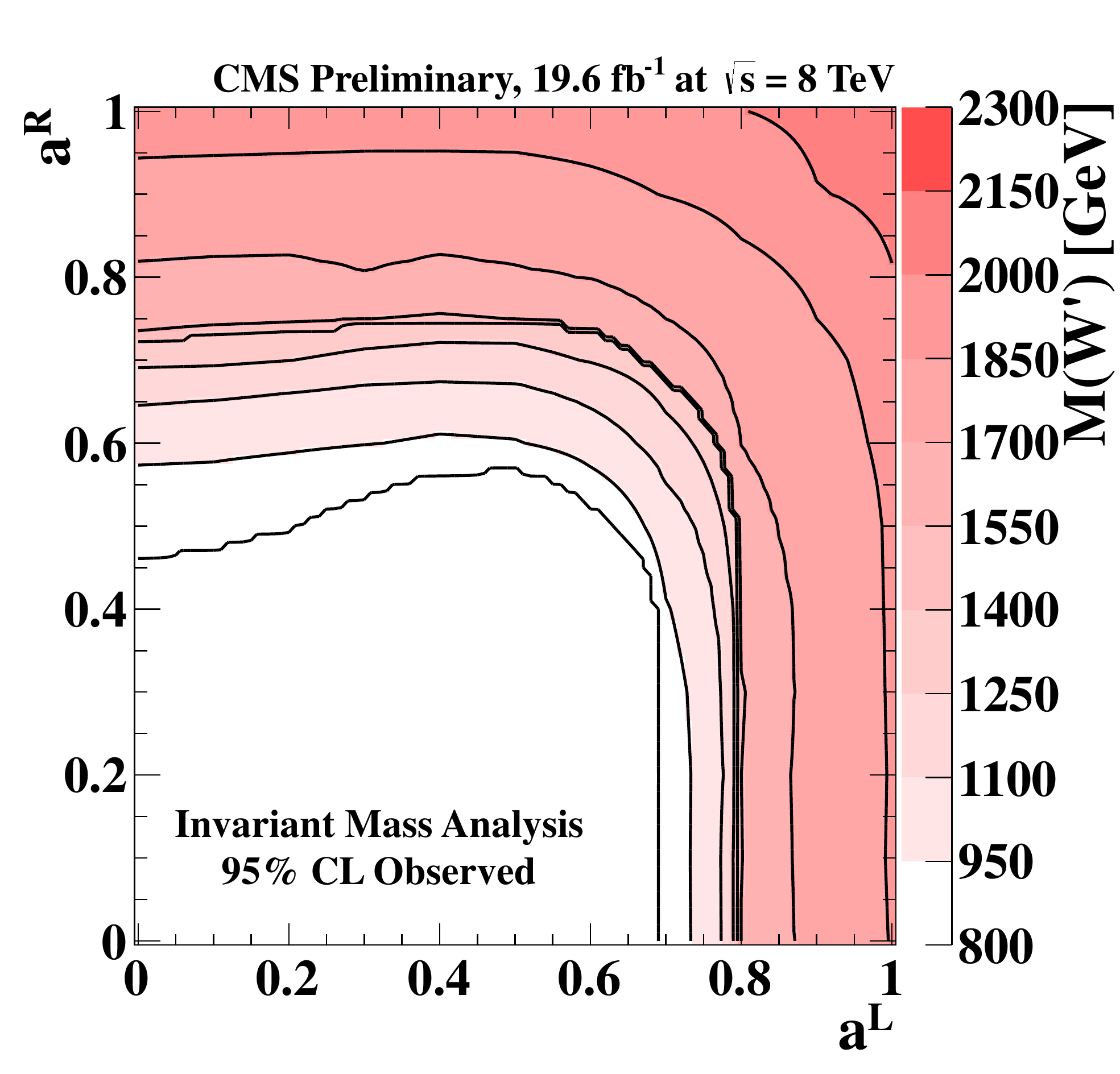} 
\end{center}
\vspace{-0.5cm}
\caption[]{
\textit{Left:} Reconstructed invariant mass distribution of selected $tb$ candidates in the $\mu$+jets channel. The expected SM backgrounds from \ttbar and single top production are shown in red and backgrounds from vector boson production in green. Simulated signal events are shown as black lines for different masses of right-handed $W_R'$ bosons, scaled by a factor of 20. \textit{Right:} Observed upper limits on the $W'$ mass in the left- and right-handed coupling plane $(a^L, a^R)$. 
\label{fig:wprime_results}
}
\end{figure}
In order to enhance the sensitivity, the mass of the $W'$ is reconstructed, using the known $W$ and top masses as constraints to find the top quark candidate.
The obtained four-vector is combined with the remaining highest \pt jet in the event to obtain the invariant mass $M_{tb}$ of the $tb$ system. Additional cuts on the \pt and invariant mass of the reconstructed top quark and the \pt of the vector sum of the two leading jets are imposed to increase the sensitivity. These requirements result in a reduction of the SM background by about 70\%, while reducing the signal selection efficiency only by 30\%. 

The simulation of the shape of the SM \ttbar and $W$+jet background processes is controlled by defining signal-depleted sideband regions. For the $W$+jet background this is achieved using a sample without $b$-tagged jets, where good agreement between the data and the $W$+jet simulation is found. using simulated $W$+jet events, it is also verified that the $M_{tb}$ distribution is independent of the number of $b$-quarks produced. 
The \ttbar simulation is verified in a control sample 
which ensures small signal contamination and large suppression of the $W$+jet background. 
It is found that the \pt of the top quark is not simulated accurately. A correction is derived which is applied to all \ttbar events. The difference between the corrected and original \ttbar simulation is taken as systematic uncertainty.

The resulting distribution of $M_{tb}$ is shown in Fig.~\ref{fig:wprime_results} (left) for the $\mu$+jets channel. No deviation from the expected SM background is observed with a similar result in the $e$+jets channel. Both channels are combined to set upper limits on the $W'$ cross section, which can be translated in excluded mass ranges for the $W'$ boson. Contours of the 95\% confidence level lower limits for $W'$ masses are shown in Fig.~\ref{fig:wprime_results} (right) in the plane for left- and right-handed couplings to fermions $a^L$ and $a^R$. In the limit of $a^L=0$ and $a^R=1$, $W'$ bosons are excluded at 95\% confidence level for masses below 2.03~TeV (with 2.09~TeV expected). 

\vspace{-0.2cm}
\section{Resonances in $\bm{t + }$jet production}
\vspace{-0.2cm}
\begin{figure}[t!]
\begin{center}
\includegraphics[width=0.4\textwidth]{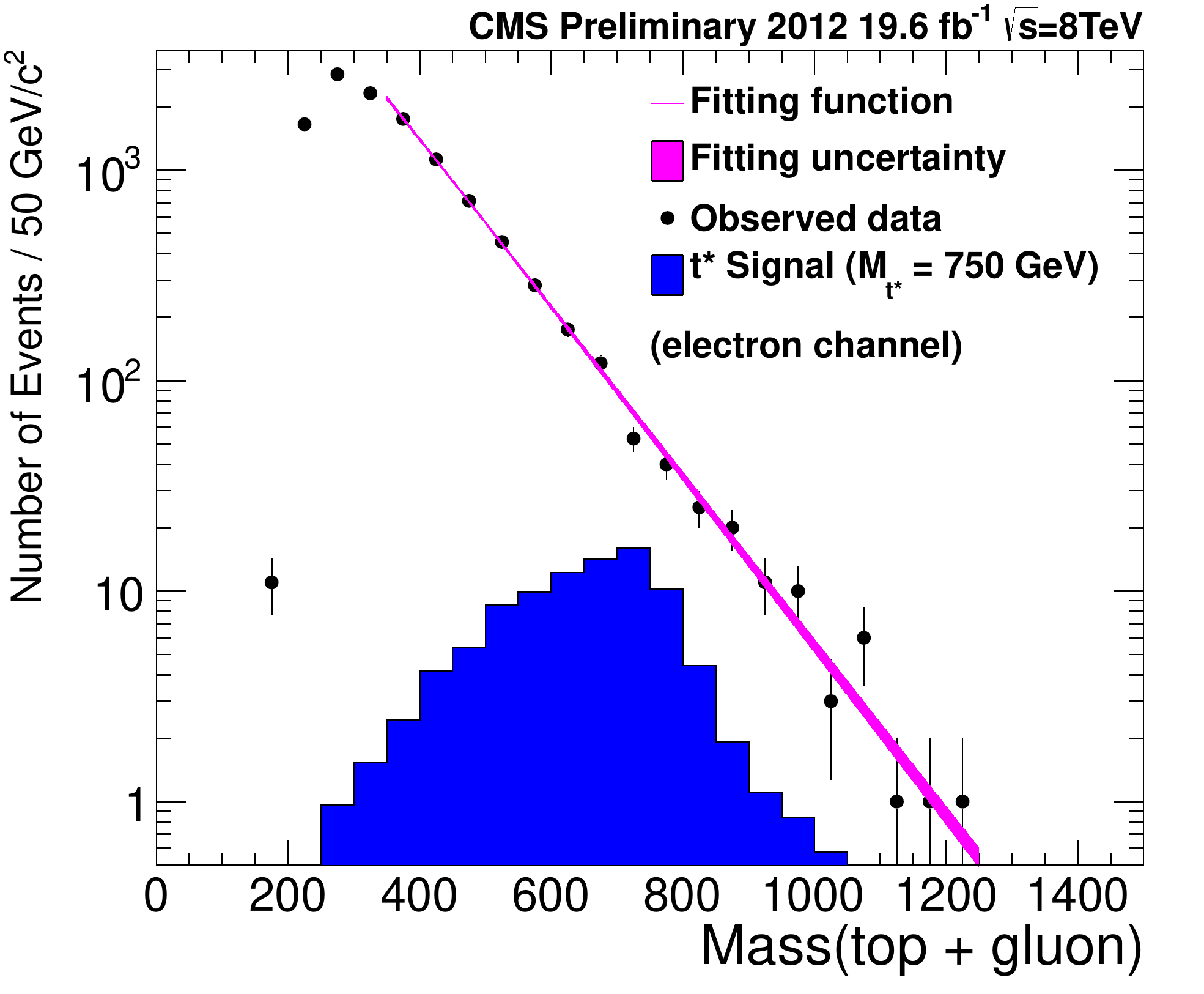} \hspace{0.2cm} 
\includegraphics[width=0.5\textwidth]{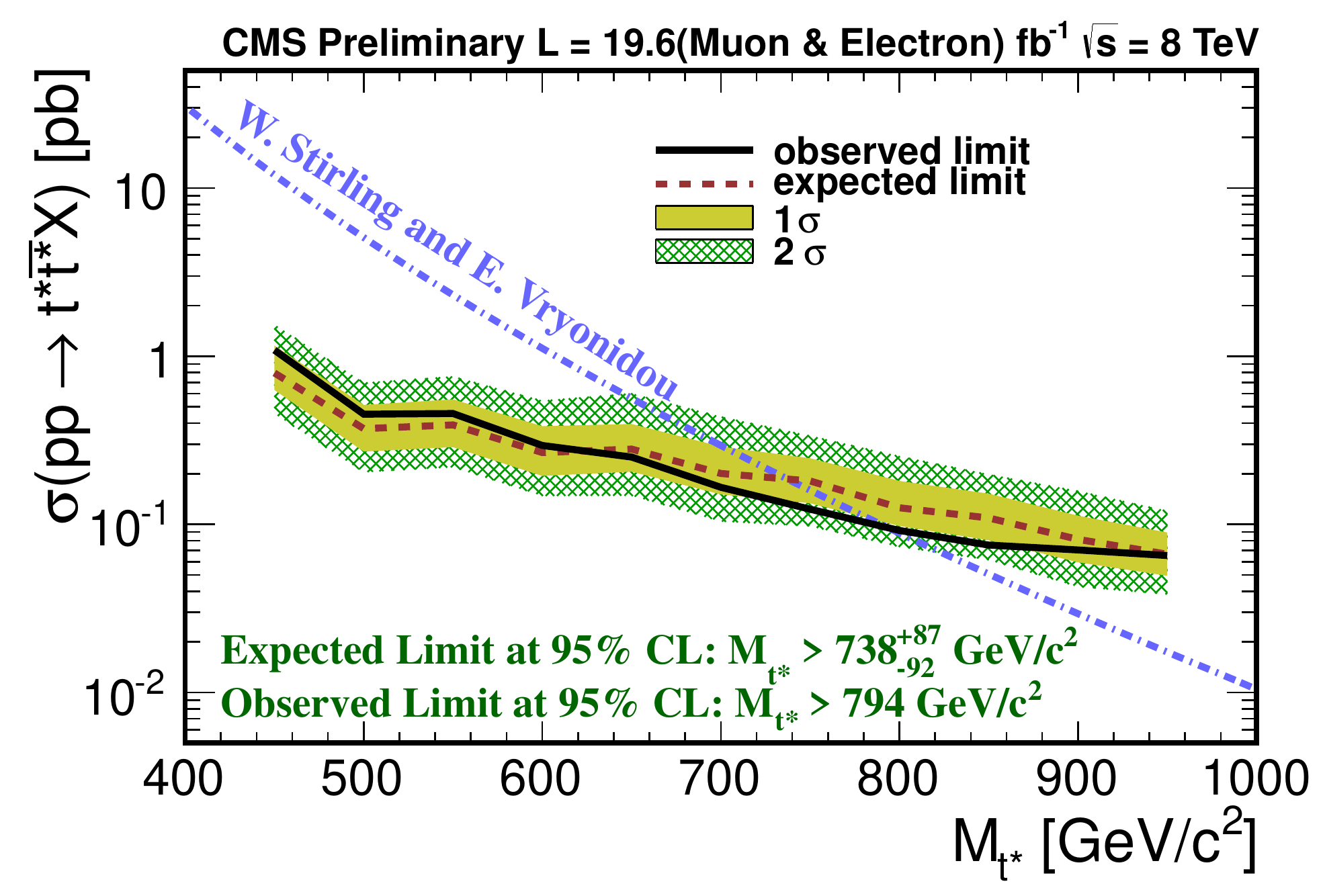} 
\end{center}
\vspace{-0.5cm}
\caption[]{
\textit{Left:} Reconstructed invariant mass distribution of the top+jet system in the electron channel. The background-only fit is shown in pink, while the expected distribution from a $t^* \to t g$ with a mass of 750~GeV is shown as solid histogram.
\textit{Right:} Observed and expected upper limits at 95\% confidence level on the production times branching ratio of $t^*$ pair production and subsequent decay $t^* \to t g$. Also shown is the predicted cross section for $t^*$ pair production.
\label{fig:tstar_results}
}
\end{figure}
The existence of an excited top quark $t^*$ is investigated in searches for resonances in the $\ttbar + 2$ jets final state, which would result from a pair-produced $t^*$ with subsequent decay $t^* \to tg$. 
A dedicated search for these resonances~\cite{B2G-12-014} selects events based on the $\ell$+jets channel requiring one isolated lepton and six or more jets, one of which must be $b$-tagged. After this selection the number of events in data is smaller than the SM expectation by about 12\%. This is consistent with recent studies of \ttbar production at high jet multiplicities~\cite{TOP-12-018} and well within the uncertainty of the SM expectation of about 30\%. 
To ensure that a potential mismodelling of the \ttbar system with additional jets does not affect this search, a strategy is chosen to determine the SM background from data. 
The \ttbar system is reconstructed with a kinematic fit and the reconstructed invariant mass distribution of the top+jet system is modelled by a Fermi-like function as shown in Fig.~\ref{fig:tstar_results} (left). Also shown is the expected distribution from $t^*$ pair production. Upper limits on the signal cross section are obtained from the maximum-likelihood fit, where the parameters of the background function and the signal normalisation are determined simultaneously. The resulting upper limits on the production cross section as function of the mass of the $t^*$ are shown in Fig.~\ref{fig:tstar_results} (right) for a combination of the electron and muon channels. The resulting lower limit for the mass of the spin-3/2 $t^*$ in the Randall-Sundrum model is 794~GeV.

\vspace{-0.2cm}
\section{Resonances in $\bm{t\bar{t}}$ production}
\vspace{-0.2cm}
The presence of additional heavy gauge bosons decaying predominantly to \ttbar would manifest itself in an enhancement in the invariant mass distribution $m_{\ttbar}$ of the \ttbar system. Several SM extensions suggest the existence of such resonances, in the following generically referred to as $Z'$. For masses $M_{Z'}$ kinematically accessible at the LHC and widths $\Gamma_{Z'}$ not too large ($\Gamma_{Z'} / M_{Z'} \lesssim 0.1$), a resonance structure would be visible in the $m_{\ttbar}$ distribution. Also, for relative widths $\Gamma_{Z'} / M_{Z'}$ less than the experimental resolution of the order of 0.1, searches for a resonance structure can be considered model-independent. For higher masses and higher width the presence of a $Z'$ boson would lead to a deviation of the \ttbar cross section from the SM prediction at high $m_{\ttbar}$ and consequently at high top quark $\pt$. 

Searches for resonances in the $m_{\ttbar}$ distribution need to be adjusted depending on the mass of the $Z'$ boson. In case of low masses, $M_{Z'} \lesssim 1$~TeV, and therefore intermediate top quark $\pt < 500$~GeV, the event selection follows those of traditional \ttbar cross section measurements.  
In case of higher masses, the decay products of the top quarks will be collimated and non-resolved topologies need to be considered in order to ensure stable selection efficiency over the accessible mass range.

A search for $Z'$ resonances by the CMS Collaboration in the $\ell$+jets channel considers both, resolved and merged, topologies~\cite{B2G-12-006}. The resolved analysis selects events with one isolated lepton, missing transverse momentum and four or more jets with $\pt>30$~GeV, where at least one has to be $b$-tagged. The two leading jets are required to have $\pt>70$~GeV and $\pt>50$~GeV to suppress background from $W$-boson and Drell-Yan production. 
In the boosted channel 
the lepton from the $W$ decay is not well separated from the $b$-quark jet anymore due to the high top quark \pt, and a selection based on non-isolated leptons is applied. In order to suppress multijet background from light-jet production a cut is applied depending on the separation in azimuth and pseudorapidity and the relative \pt between the lepton and nearest jet. This selection ensures a negligible contribution from multijet background and a signal efficiency of about 70\%. 
Since the decay products from the hadronically decaying $W$-boson may overlap and get merged with the $b$-quark jet, only two or more jets are required where the two leading jets have to fulfil the requirements $\pt>150$~GeV and $\pt>50$~GeV. 
In both kinematic regimes the \ttbar system is reconstructed by an hypothesis test which considers all possible combinations of jets with the lepton-neutrino system.
In the boosted analysis an additional cut on the discriminating value of this hypothesis test is applied, which removes a large fraction of the background from $W$+jet production. 

The two analyses use complementary methods to estimate the SM background. In the threshold analysis a maximum-likelihood fit to the falling $m_{\ttbar}$ distribution is performed. This fit determines the background function and normalisation of the signal templates simultaneously. 
The boosted analysis uses predictions from MC simulations for the background estimation. 

\begin{figure}[t!]
\begin{center}
\includegraphics[width=0.47\textwidth]{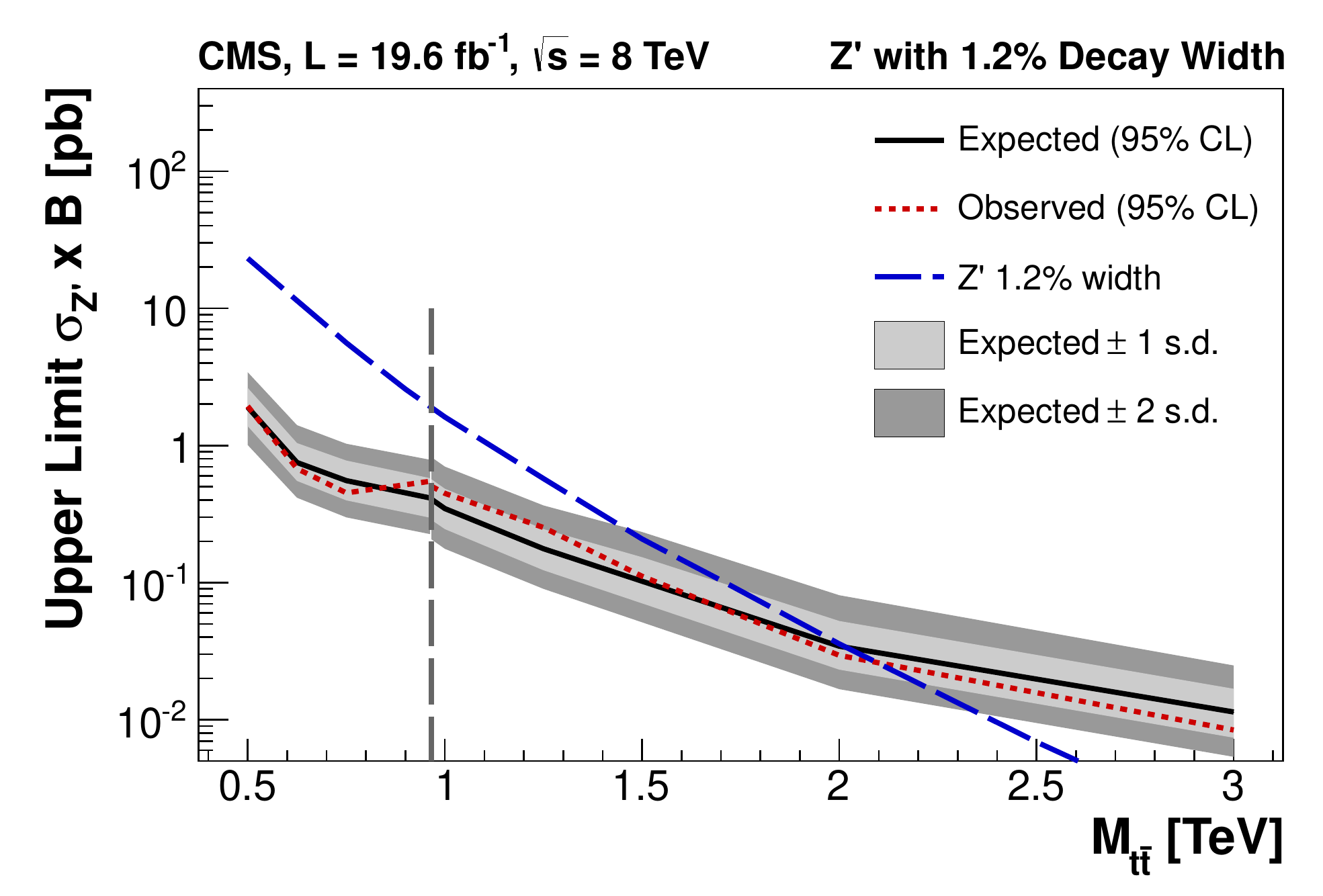} 
\raisebox{2mm}{
\includegraphics[trim=5mm 5mm 2mm 0mm, clip, width=0.48\textwidth]{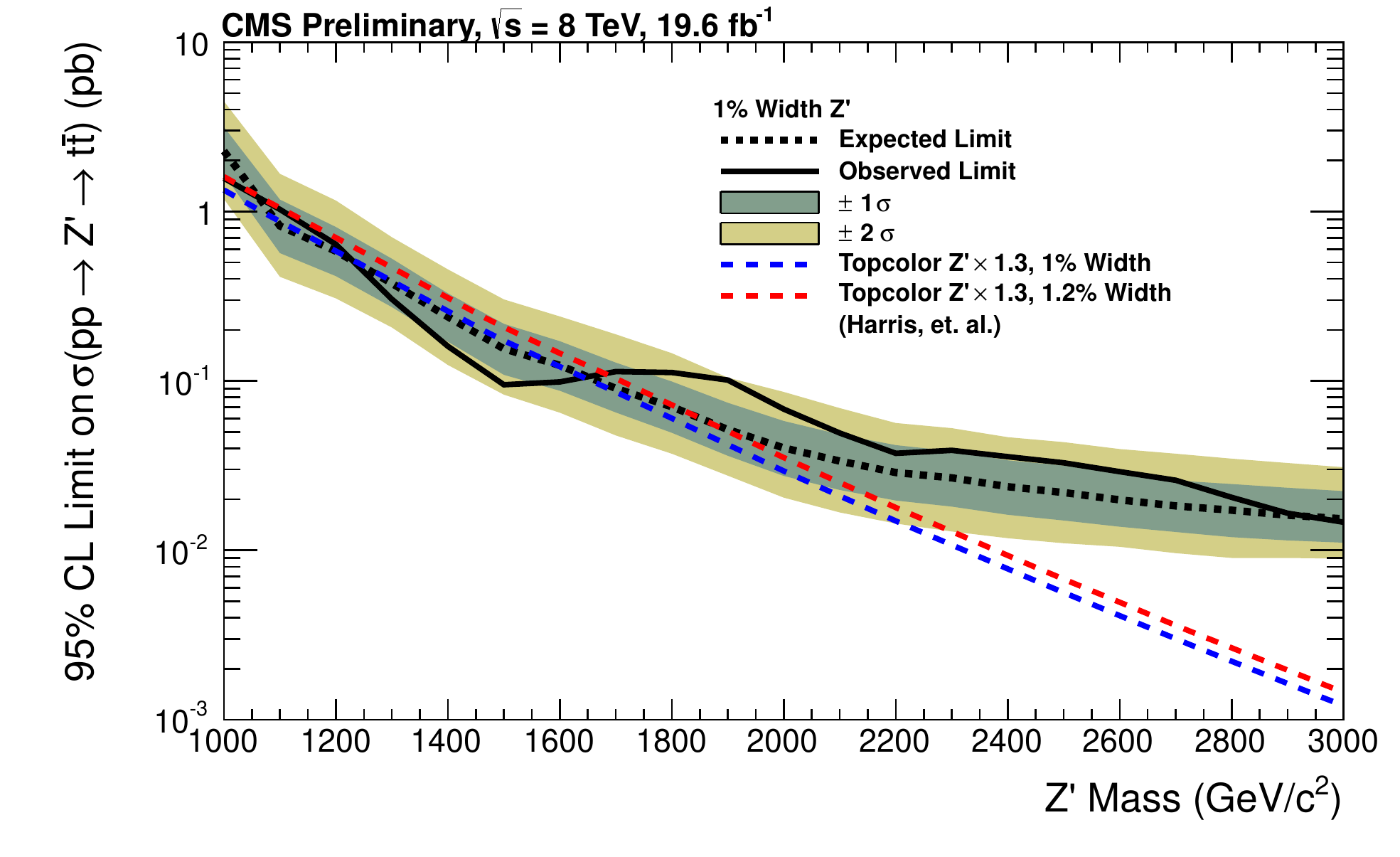} 
}
\end{center}
\vspace{-0.5cm}
\caption[]{
Observed and expected upper limits at 95\% confidence level on the production times branching ratio of $Z' \to \ttbar$ in the $\ell$+jets channel (\textit{left}) and in the all-hadronic channel (\textit{right}). Also shown is the predicted cross section for $Z'$ production with relative width of 1.2\%.
\label{fig:zprime_results}
}
\end{figure}
The combined results from the threshold and boosted analyses, expressed as upper limits on the production cross section times branching ratio of $Z' \to \ttbar$, are shown in Fig.~\ref{fig:zprime_results} (left) for resonances with 1.2\% relative width. The vertical dashed line indicates the transition point between the two analyses, which is based on the sensitivity of the expected limits. For $Z'$ masses above about 1~GeV the boosted analysis has higher sensitivity. A leptophobic topcolor $Z'$ model~\cite{Harris:2011ez} can be excluded for masses below 2.1~TeV. 

A complementary search for \ttbar resonances is performed in the all-hadronic channel~\cite{B2G-12-005}, which is optimised for sensitivity at high masses ($M_{Z'}\gtrsim 1$~TeV). 
The collimated decay products of the top quarks result in a dijet signature. A top-tagging algorithm based on Ref.~\cite{Kaplan:2008ie} is used to separate signal events from multijet production of light-quark jets. The event selection is based on a back-to-back dijet signature requiring two jets with $\pt > 400$~GeV and an azimuthal separation of $\Delta\phi > \pi / 2$. Events are only accepted if both jets fulfil the top-tagging requirements of having at least three subjets, a jet mass of $140 < m_{\rm jet} < 250$~GeV and a minimum pair-wise mass of the subjets $m_{\rm min}> 50$~GeV. 
After the final selection about 93\% of the background originates from multijet production, with only 7\% irreducible background from \ttbar production. While the \ttbar background is taken from simulation, the multijet background is derived from data. This is done by measuring the misidentification rate of the top-tagging algorithm in a signal-depleted sideband region, by reversing the requirement on $m_{\rm min}$ on one jet and measuring the misidentification rate depending on \pt on the other jet. 
This mistag rate is used to predict the background from multijet production in the signal region.

The results obtained in this analysis are shown in Fig.~\ref{fig:zprime_results} (right). While the sensitivity is worse at $M_{Z'}\sim1$~TeV than in the $\ell$+jets analysis, at resonance masses of around 2 TeV the two analyses give comparable expected limits. 
Further improvements on the sensitivity can be achieved by a statistical combination of both analyses, which has been published recently~\cite{Chatrchyan:2013lca}.


\end{document}